\documentclass[fleqn,usenatbib]{mnras}
\usepackage{newtxtext,newtxmath}
\usepackage[T1]{fontenc}

\DeclareRobustCommand{\VAN}[3]{#2}
\let\VANthebibliography\thebibliography
\def\thebibliography{\DeclareRobustCommand{\VAN}[3]{##3}\VANthebibliography}

\usepackage{color}
\usepackage{enumitem}
\usepackage{hyperref}
\usepackage{verbatim}
\usepackage{amsmath,amstext,mathrsfs}
\usepackage[all]{hypcap} 
\usepackage{afterpage}    
\usepackage{marginnote}
\usepackage{makecell}
\usepackage{subfigure}
\usepackage{xfrac}

\usepackage{ifxetex,ifluatex}
\usepackage{fixltx2e} % provides \textsubscript
\usepackage{multirow}

\DeclareFontFamily{OMS}{oasy}{\skewchar\font48 }
\DeclareFontShape{OMS}{oasy}{m}{n}{%
         <-5.5> oasy5     <5.5-6.5> oasy6
      <6.5-7.5> oasy7     <7.5-8.5> oasy8
      <8.5-9.5> oasy9     <9.5->  oasy10
      }{}
\DeclareFontShape{OMS}{oasy}{b}{n}{%
       <-6> oabsy5
      <6-8> oabsy7
      <8->  oabsy10
      }{}
\DeclareSymbolFont{oasy}{OMS}{oasy}{m}{n}
\SetSymbolFont{oasy}{bold}{OMS}{oasy}{b}{n}

\DeclareMathSymbol{\smallleftarrow}     {\mathrel}{oasy}{"20}
\DeclareMathSymbol{\smallrightarrow}    {\mathrel}{oasy}{"21}
\DeclareMathSymbol{\smallleftrightarrow}{\mathrel}{oasy}{"24}
 
\newcommand*{\torefereetwo}{}
\newcommand*{\toreferee}{ }
\graphicspath{{./}{Fig/}}

\date{Accepted XXX. Received YYY; in original form ZZZ}

\begin{document}

\title[]{Tracing of Magnetic field with gradients: Sub-Sonic Turbulence}

\author[Ho \& Lazarian]{
K. W. Ho, $^{1}$ \thanks{E-mail: kho33@wisc.edu}
A. Lazarian, $^{1,2}$ \thanks{E-mail:  alazarian@facstaff.wisc.edu}
\\
$^{1}$Department of Astronomy, University of Wisconsin-Madison, Madison, WI, 53706, USA\\
$^{2}$Centro de Investigación en Astronomía, Universidad Bernardo O’Higgins, Santiago, General Gana 1760, 8370993, Chile\\
}

\date{Accepted 2023 January 12. Received 2023 January 12; in original form 2022 March 21}

\maketitle
\begin{abstract}
Recent development of the velocity gradient technique shows the {\toreferee capability} of the technique in the way of tracing magnetic fields morphology in diffuse interstellar gas and molecular clouds. In this paper, we perform the numerical systemic study of the performance of velocity and synchrotron gradient for a wide range of magnetization in the sub-sonic environment. Addressing the studies of magnetic field in atomic hydrogen, we also study the formation of velocity caustics in the spectroscopic channel maps in the presence of the thermal broadening. We show that the velocity caustics can be recovered when applied to the Cold Neutral Medium (CNM) and the Gradient Technique (GT) can reliably trace magnetic fields there.  Finally, we discuss the changes of the anisotropy of observed structure functions when we apply to the analysis the procedures developed within the framework of GT studies.
\end{abstract}


\section{Introduction} \label{sec:intro}
Magnetic fields are very important for key astrophysical processes in interstellar media (ISM) such as the formation of stars (see \citealt{MO07,MK04}), the propagation and acceleration of cosmic rays (see \citealt{J66,YL08}), the regulation of heat and mass transfer between different ISM phases (see \citealt{Draine2009} for the list of the different ISM phases). Polarized radiation arising from the presence of the magnetic field also interferes with the sygnal of the enigmatic CMB B-modes arising from gravity waves in the early Universe. \citep{1997PhRvD..55.1830Z,2017PhRvL.118i1801C,Kandel17}. Therefore, it is essential to have a reliable way to study the properties of magnetic fields in those process.

The traditional way to study the Plane of Sky (POS) magnetic fields is using polarimetry measurements \citep{Planck,L07}. It is widely used from radio to optical wavelengths to trace the magnetic field morphology at various scales in the ISM.

Recently, a new promising technique has been proposed, the velocity gradient technique (VGT), which is capable of tracing magnetic field using spectroscopic data \citep{YL17a,LYH18,HuNature,HL20a}. The technique makes use of the fact that magnetic fields make turbulence anisotropic, with turbulent eddies being elongated along the magnetic field (See \cite{BL19} for a monograph). As a result, the turbulence induces the fluid motion mostly perpendicular to the direction surrounding magnetic eddies. It is important that the magnetic field direction is the {\it local} direction of magnetic field in the vicinity of turbulent eddies. This follows directly from the theory of turbulent reconnection that predicts that magnetic fields of the eddies reconnect over one eddy turnover time (\cite{LV99}, hereafter LV99). This property of magnetic turbulence is central for magnetic field tracing with both velocity gradients as well as other types of gradients, e.g. synchrotron intensity gradients \citep{LY17a}, synchrotron polarization gradients \citep{LYH18}.

The VGT has been numerically tested for a wide range of column densities from diffuse transparent gas to molecular self-absorbing dense gas \citep{YL17a, LY18a, HuNature, HL21}. The technique was shown to be able to provide both the orientations of the magnetic field as well as a measure of media magnetization \citep{LYH18}.  A VGT survey was conducted recently to study the morphology of a few nearby molecular cloud \citep{HuNature}. {\toreferee The result showed consistency with the Planck polarization measurement and indicate the capability of the VGT on 
tracing magnetic field in different ISM region. }

While the earlier VGT study mainly focused on the supersonic spectroscopic data, the same idea of tracing magnetic with gradients can be employed with different types of astrophysical data. For instance, \cite{LY17a} showed gradient can also be applied to trace magnetic field with  synchrotron intensity gradients (SIGs) maps. The corresponding emission comes from subsonic warm/hot media. The lack of shock wave in sub sonic environment is beneficial for magnetic field tracing. 

Tracing of magnetic field in subsonic media is also important within the VGT. The velocity gradients can be obtained in this setting using velocity centroids which are not sensitive to thermal broadening. If the channel maps are applied to subsonic data, first of all, one can use heavier species as spectroscopic tracers. For such species, the thermal broadening is suppressed and caustics produced in channel maps are prominent. In addition, the newly introduced Velocity Decomposition Algorithm (VDA) \cite{Yuen2021a} opens ways of exploring velocity caustics in the presence of the thermal broadening. 

Therefore this study explores the ability of magnetic field tracing using both the VGT and the SIGs for subsonic medium. Several concerns arise on the application of Gradient Technique (GT) in the sub-sonic environment. First, multi-phase media study (see \cite{Yuen2021a}) shows that thermal broadening is a crucial factor that smooths out the structure in the subsonic spectroscopic data. It may potentially weaken the ability of the VGT to trace the magnetic field. Second, \cite{HL20a} found out that the intermittency of fast mode could also play an important role in affecting the VGT analysis. In the case that the fast mode dominate the energetics of a particular region they induce there the rotation of the velocity gradient direction from parallel to perpendicular to the magnetic field. This, however, does not happen with the SIGs, for which the gradients induced by fast and Alfven modes are parallel. 

In \cite{HL20a} we proposed a new technique, Gradient of Gradient Amplitude (GGA),  which improves the magnetic field tracing by gradients. However, an in-depth study is required to analyze the applicability of GGA in sub-sonic regime versus the change of {\toreferee Alfven Mach number}.

Below, we perform a new study of the GT in the sub-sonic environment to answer the concerns above and evaluated the performance of the GT in a low Ms regime. 
In what follows, we would cover the theory in section \ref{sec:Theory} and our numerical setup in section \ref{sec:Methodology}. Then we would discuss the result of the alignment measure of the gradient in the ideal observable measure and the velocity gradient in the presence of thermal broadening in multi-phase media in section \ref{sec:Result}. We further extend the study of GGA in section \ref{sec:GAG_imporvement}. We then discuss the the Correlation Function Analysis (Hereafter CFA) alignment in section \ref{sec:CFA}. At last, we would discuss our work in section \ref{sec:discussion} and summarize the paper in section \ref{sec:conclusion}.

% Magnetohydrodynamic

\section{Gradient Technique} \label{sec:Theory}
\subsection{Theoretical Considerations}

The most important component of {\toreferee Magnetohydrodynamic (MHD)} turbulence is the cascade of Alfvenic motions. Therefore, below we will focus on the properties of Alfven modes.

The modern theory of MHD turbulence originates from the work of \citealt{GS95} (henceforth GS95) that described the scaling of transAlfvenic  incompressible turbulence in what is now known to be the strong MHD turbulence regime. The description was, however in the frame of the mean magnetic field, which, as it was shown by the later studies, the GS95 statitical scalings are not applicable. 

Further advances were related to understanding of the importance of the {\it local} system of reference as well as the generalization of the theory for the sub-Alfvenic regime in \citealt{LV99} (henceforth LV99). There also the regime of weak turbulence was quantified (see also \cite{Galtier2000}). 

The local system of reference is the system of reference in respect to which the turbulent motions should be considered. Its importance is easiest to see considering magnetic eddies. Due to fast turbulent reconnection the eddies aligned with the magnetic field direction in their vicinity can reconnect and perform a turnover within one eddy turnover time (LV99). This happens on the eddy turnover scale $\sim l_\bot/v_l$, {\toreferee where $l_\bot, v_l$ are the size of eddy perpendicular to the local magnetic field direction and the eddy's velocity at the scale l}. Incidentally, this mixing results in inducing an Alfven perturbation with the same period, i.e. $l_\bot/v_l\sim l_\|/V_A$, where $V_A$ is the Alfven velocity. The latter corresponds to the condition termed "critical balance" in GS95. However, unlike the origianal GS95 claim, the critical balance is only in the system of reference aligned with the local direction of the magnetic field, i.e. with the direction of the magnetic field in the direct vicinity of the eddy. The local system of reference is absolutely critical for the GT. It is only because of the localized alignment that the gradients of velocity and magnetic field can trace 3D magnetic field.

The numerical study in \citealt{CV00,MG00} established numerically the vital importance of the local system of reference for the description of MHD turbulence. The subsequent studies in \cite{LG01} as well as \cite{CL02,CL03,Kowal09}, extended the the theory to the compressible case. This theory of MHD compressible turbulence (see the monograph by \cite{BL19}) is at the basis of the GT. 

It is important to note that the motions perpendicular to the local magnetic field have the form of Alfvenic eddies and they exhibit Komlogorov scaling $v_l\sim l_\bot^{1/3}$. Therefore the gradients scale as $v_l/l_\bot\sim l_\bot^{-2/3}$, meaning that the gradients at the smallest resolved scales are the most important (see \cite{L20} for the analytical theory of gradient measurements). These gradients are perpendicular to the magnetic field and their direction should be turned 90 degrees to get the magnetic field tracing. It is important that the amplitude of the gradients increases with the decrease of the scale. Therefore, the gradients measured at the smallest scales are the most prominent. These gradients, similar to aligned grains (see \citep{Andersson2015}), sample the 3D magnetic field along the line of sight. Due to this effect, the large scale gradients, e.g. arising from galactic shear, are not important for the analysis of the high resolution data.

%why gradient is an universal approach ? underlying physics is the same.

\subsection{Velocity and magnetic gradients}

\subsubsection{General outlook}

The 3D velocity fluctuation are not directly available from the observations. Instead, the gradients of velocity centroids and the gradients of intensity fluctuations measured within thin channel maps \footnote{{\toreferee For a channel maps with channel width $\Delta v$, the thin channel map means its  $\Delta v \leq \sqrt{\delta v^2_R}$, where $\delta v_R$ is the velocity dispersion. }} can be used as proxies of the velocity gradients.{\toreferee In both cases, the gradients are measured for turbulent volume extended by $\mathcal{L}>L_{inj}$ along the LOS, and this entails additional complications, where $\mathcal{L}, L_{inj}$is the LOS depth and the injection scale}. While eddies stay aligned with respect to the local magnetic field, the direction of the local magnetic field is expected to change along the LOS. Thus, the contribution of 3D velocity gradient are also summed up along the line of sight.

The spectrum of observed fluctuations changes due to the averaging effect along the LOS. It is easy to show that the 2D spectrum of the turbulence obtained by projecting the fluctuations from 3D has the same spectral index of -11/3 \footnote{{\toreferee Starting from 1D spectrum $P_{1D}$ with spectral index -5/3, we can get back 3D spectral index of $-11/3$ by considering the dimensional analysis of $P_{3D}$ = $P_{1D} k^{-2}$}}.  The relation between the spectral slope of the correlation function and the slope of the turbulence power spectrum in 2D in this situation is $- 11/3+2 = - 5/3$, where 2 is the dimensionality of the space. Therefore, the 2D velocity fluctuations arise from the 3D Kolmogorove-type turbulence scale as $l^{5/6}_{2D}$ with the gradient anisotropy scaling as $l^{-1/3}_{2D}$. It is important that the amplitude of the gradients increases with the decrease of the scale. Therefore, the gradients measured at the smallest scales are the most prominent. These gradients, similar to aligned grains (see \citep{Andersson2015}), sample the 3D magnetic field along the line of sight. Due to this effect, the large scale gradients, e.g. arising from galactic shear, are not important for the analysis of the high resolution data.

The slow modes follow the scaling of the Alfven modes \citep{GS95,LG01,CL02,CL03} and therefore induce the same type of gradients as Alfvenic modes. while fast modes are different \citep{CL02,CL03,Kowal09,HL20a}). It follows from the theory in (\cite{LP12}, hereafter LP12) that gradients of synchrotron emission arising from fast modes are also aligned perpendicular magnetic field direction, while the anisotropies of the gradients of velocity caustics and velocity centroids are different \citep{Kandel17,Kandel18}. It is possible to show \citep{LYH18} that the corresponding gradients are perpendicular to those created by Alfven and slow modes. Therefore, the contribution of the fast modes can decrease the accuracy of the GT. We are dealing with their contribution in this paper.

\subsubsection{VGT for molecular clouds and diffuse HI}

The magnetic field tracing with velocity gradients in molecular clouds can be tested successfully with isothermal numerical simulations (see \cite{HuNature}). This is due to efficient cooling of the molecular clouds, which is different from HI gas (See \cite{Field1969,Wolfire1995,Wolfire2003}). The HI gas is stabilized by the thermal equilibrium between the heating and cooling and forms two stable phases: the warm and cold phases. Other than the two phases, the thermally unstable phase also plays a vital role in the atomic hydrogen environment due to the consequence of strong turbulence. Due to the presence of magnetized turbulence in the atomic hydrogen it is a promising medium of applying the VGT. In such an environment, the VGT has already demonstrated the reliable tracing of the magnetic field \citep{YL17a,HuNature}.

The turbulence is subsonic in most volume of galactic HI, which corresponds to the warm phase.\citep{Saury14,Marchal21} The Velocity Decomposition Algorithm (VDA) developed in \cite{Yuen2021a} allows to identify velocity caustics produced in this phase.
{\toreferee 
\subsection{Velocity caustics}
The concept of velocity caustics is first proposed by \cite{LP00} and further facilitated by \cite{Yuen2021a}. Velocity caustics describes the effect of pure turbulent velocity fluctuation and how they come into the thin channel map. One ideal picture would be, even though considering a incompressible magnetized turbulent fluid with no density fluctuation, we can still observe a channel map with anisotropic fluctuation arising from the turbulence. {\torefereetwo Those fluctuations are often referred to as the velocity contribution} and different statistical tools (for example, VGT) could {\torefereetwo utilize} the information to trace magnetic field. {\torefereetwo However, the fluid} contains compressibility and density contamination caused by thermal broadening effect, making the fluctuation of channel map contains the contribution from both density and velocity part. Nonetheless, the density effect on sub-sonic media is sub-dominate and can be removed by using the algorithm proposed by \cite{Yuen2021a}.
}
\subsubsection{Synchrotron emission}

Measurements of polarized synchrotron radiation and Faraday rotation (see \cite{2013pss5.book..641B,2015A&A...575A.118O,2011MNRAS.412.2396F,2016ApJ...830...38L,2017A&A...597A..98V} ) provide an important insight into the magnetic structure of the Milky Way and the neighboring galaxies. Synchrotron radiation fluctuation carries the statistical information of MHD turbulence.  Serial studies discussed how to apply gradient onto measurable quantities, such as synchrotron intensity and synchrotron polarization (See \cite{LY17a,LY18a}). In this paper we focus on the gradient on synchrotron intensity map as it is a observable that we  deal with. 

For the power-law distribution of electrons \( N(E) E \sim E^\alpha dE \), the synchrotron emissivity is
\begin{equation}
\begin{aligned}
I_{sync}({\bf X}) \propto \int dz B^{\gamma}_{POS} (\bf X,z)
\end{aligned}
 \end{equation}
where $B^{\gamma}_{POS} = \sqrt{B_x^2+B_y^2}$ corresponds to the magnetic field component perpendicular to the {\toreferee line of sight}, ${\bf X}$ is the plane of sky vector defined in x and y direction, z the line of sight axis and, $B_x,B_y$ the 3D magnetic field in x and y direction. The fractional power of the index $\gamma = (\alpha+1)/2 $ was a impediment for quantitative synchrotorn statistical studies. However, LP12 showed that the correlation functions and spectra of the $B^{\gamma}_{\bot}$ could express as $\alpha = 3$, which gives $\gamma$ and therefore the dependence of synchrotron intensity on the squared magnetic field strength.

\subsection{Application of Gradient in Sub-Sonic Environment}
Below we will discuss two important examples to which we will apply the GT. Those are the centroid map and the synchrotron intensity map. We will perform a systematic study of the GT by changing the magnetization of the numerical data used to produce synthetic observations.  Other than that, we would also like to study the behavior of GT in the HI spectroscopic velocity channel maps due to the recent debate of the velocity caustics effect in the channel map (See section 4.3 and 7.2 for more information). 

\section{Numerical Simulation and Measures employed}
\label{sec:Methodology}
\subsection{Simulation Setup}
The numerical data that we analyzed in this work are obtained by 3D MHD simulations using the single-fluid, operator-split, staggered-grid MHD Eulerian code ZEUS-MP/ 3D \citep{Hayes06} to set up a 3D, uniform, and isothermal turbulent medium. Periodic boundary conditions are applied to emulate a part of the interstellar cloud. Solenoidal turbulence injections are employed.  To extend our study from super sonic regime to sub sonic regime, we simulate two sets of ensemble in each regime. Two sets of simulations employ various Alfvenic Mach numbers $M_A = V_L/V_A$ with Sonic Mach Number $M_S = V_L/V_S$ at about 6 and 0.5 where $V_L$ represents the injection velocity, $V_A$ the Alfven velocities, $V_s$ the sonic velocity. For the generation of turbulence, the turbulence is injected solenoidally for all the simulations using the Fourier-space method. Turbulent energy is injected at the large scale ( k=2 ) and {\torefereetwo dissipated by the viscosity at small scale.} We adjust the strength of the injection such that {\torefereetwo the cubes  reach desired $M_s$ value}. All of the cubes are listed in Table 1. However, limited by the turbulence scaling (Please see LV99), we devote most of our research to the sub-Alfvenic and trans-Alfvenic case in this study. 
\begin{table}[]
\begin{tabular}{lll|lll}
\multicolumn{3}{c}{Subsonic} & \multicolumn{3}{c}{Supersonic} \\
Model    & $M_S$   & $M_A$   & Model    & $M_S$    & $M_A$    \\
H1S      & 0.67    & 0.13    & H1       & 7.31     & 0.22     \\
H2S      & 0.64    & 0.38    & H2       & 6.10     & 0.42     \\
H3S      & 0.62    & 0.64    & H3       & 6.47     & 0.61     \\
H4S      & 0.61    & 0.90    & H4       & 6.14     & 0.82     \\
H5S      & 0.61    & 1.17    & H5       & 6.03     & 1.01     \\
         &         &         & H6       & 6.02     & 1.21    
\end{tabular}
\caption{Simulation parameters where $M_S,M_A$ represents the sonic Mach number and Alfvenic Mach number. For all simulations, the resolution is set to $792^3$. $M_S,M_A$ are the sonic Mach number and the Alfvenic Mach number.}
\end{table}

\subsection{Plane of sky magnetic field}

We trace the plane of sky (POS) magnetic field orientation with polarization. We shall assume a constant-emissivity dust grain alignment process. As a comparison to gradient, we generate polarization maps by projecting our data cubes along the z-axis. We construct an synthetic Stokes parameters Q, U. 

{\torefereetwo By assuming that the constant emissivity and the dust followed the gas, which the dust uniformly aligned with respect to the magnetic field}, the Stokes parameter $Q({\bf X}),U({\bf X})$ can than be expressed as a function of angle $\theta$ at plane of sky magnetic field by $\tan(x,y) = B_y(x,y)/B_x(x,y)$ :
\begin{equation}
\begin{aligned}
Q({\bf X}, z) \propto \int dz\: \rho({\bf X},z)cos(2\theta({\bf X},z) {\toreferee )}\\
U({\bf X}, z) \propto \int dz\: \rho({\bf X},z)sin(2\theta({\bf X},z)  {\toreferee )},
\end{aligned}
 \end{equation}
where $\rho$ is the density, ${\bf X}$ is the plane of sky vector defined in x and y direction, z the line of sight axis and, $B_x,B_y$ the 3D magnetic field in x and y direction. The dust polarized intensity $I_P = \sqrt{Q^2+U^2}$and angle $\theta_p = 0.5 atan2(U/Q)$ are then defined correspondingly.

\subsection{Synchrotron intensity map}
For our present paper, we follow the approach in LP12 that amplitudes of Stokes parameters are scaled up with respect to the cosmic-ray index and the spatial variations of the Stokes parameters are similar to the case of cosmic-ray index $\gamma = 2$ . 

\subsection{Alignment Measure (AM) and sub-block averaging} To quantify how good two vector fields are aligned, we employ the {\it alignment measure} that is introduced in analogy with the grain alignment studies (see \citealt{L07}):
\begin{equation}
\label{eq:AM}
AM=2\langle\cos^2\theta_r\rangle-1,
\end{equation}
as discussed for the VGT in \citealt{GL17,YL17a}).  The range of AM is $[-1,1]$ measuring the relative alignment between the {\it $90^o$-rotated} gradients and magnetic fields, where $\theta_r$ is the relative angle between the two vectors. A perfect alignment gives $AM=1$,  whereas random orientations generate $AM=0$ {\toreferee and a perfect perpendicular alignment case refers to $AM=-1$} . In what follows we use $AM$ to quantify the alignments of VGT in respect to magnetic field.

We adopt the sub-block averaging introduced in \cite{YL17a}. {\toreferee
The use of sub-block averaging comes from the fact that the orientation of turbulent eddies with respect to the local magnetic field is a statistical concept. In real space the individual gradient vectors are not necessarily required to have any relation to the local magnetic field direction. \cite{YL17a} reported that the velocity gradient orientations in a sub-region–or sub-block–would form a Gaussian distribution in which the peak of the Gaussian fit reflects the statistical most probable magnetic field orientation in this sub–block. As the area of the sampled region increases, the precision of the magnetic field traced through the use of Gaussian block fit becomes more and more accurate. We will discuss it more in section \ref{sec:GAG_imporvement}.}

\begin{figure*}
\includegraphics[width=0.8\textwidth]{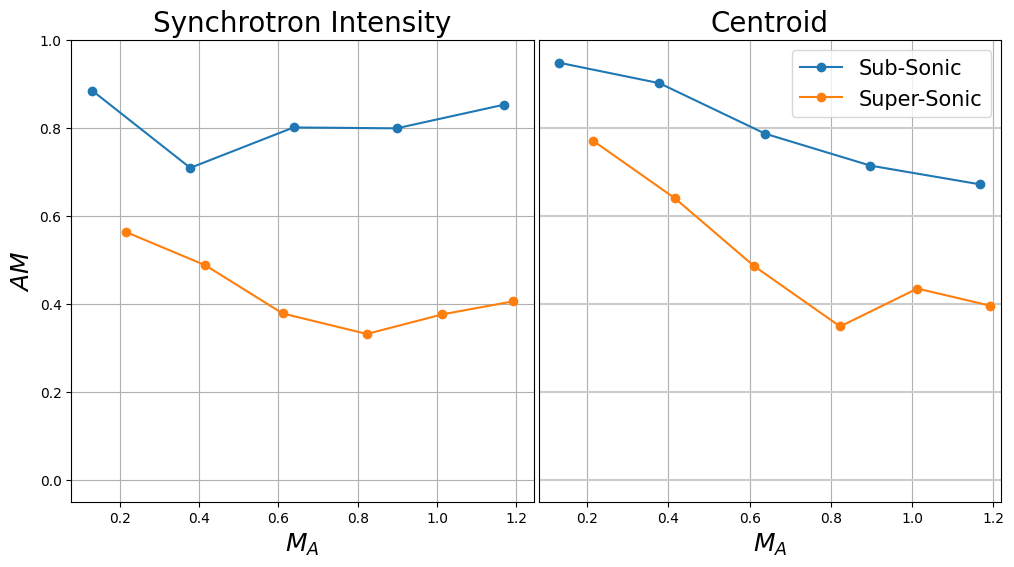}
\caption{Left panel: Result of Synchrotron intensity gradient . Right panel: Result of Centroid Gradient. Both block size used = 72.  X-axis: Alfvenic Mach Number $M_A$, y-axis: AM. The blue lines represent the AM of sub-sonic ensembles and orange lines represent the change of AM of super-sonic ensembles. }
\label{fig:SC_AM}
\end{figure*}

\section{Results}
\label{sec:Result}
For observational tracing of the magnetic field, it is essential to know what to expect in terms of AM dependence on magnetization when we employ the gradient method in the ideal synthetic environment.  We investigate how the change in Alfvenic Mach number $M_A$ would alter the tracing power of Gradient Technique (GT) with two types of data: spectroscopic maps  and synchrotron intensity map.

\subsection{Gradients of Synchrotron Intensity}
The synchrotron intensity gradient (SIG) results are presented in the left panel of Figure \ref{fig:SC_AM}. We adopt the sub-block averaging approach, and the results are computed using the block size of $72^2$. To compare the change of tracing power of GT in different hydro-dynamical regimes, we include the result of supersonic simulation($M_S \sim 6 $) with similar coverage of $M_A$ as a reference. The setting of block size is the same as the sub-sonic regime.

Throughout the change of $M_A$, the tracing power of SIG shows a different trend in different hydro-dynamical regimes.
The result of sub-sonic environments (Blue curve) shows that the tracing power of SIG is insensitive to the change of magnetization. The AM maintains at about 0.8 with a mild drop in $M_A \sim 0.4$ case. For the case of supersonic, we observe a steady downtrend of $AM$ in the sub-Alfvenic regime. The $AM$ starts at $\sim 0.58$ at $M_A\sim0.2$ and drops gradually to 0.38 at $M_A\sim0.8$. The declining trend disappears at the trans-Alfvenic and super-Alfvenic regime, which the AM steady at around 0.38. Besides, we notice that the AM of SIG in sub-sonic ensembles always higher than supersonic ensembles.

\subsection{Result of Gradient in Centroid}
For the benchmark of Velocity centroid gradient (VCG) in the sub-sonic environment, Figure. \ref{fig:SC_AM} showed the change of AM of centroid as a function of $M_A$ in the right panel. The sub-block setting is the same as SIG. As a reference, we also add the change of AM for the supersonic environment in orange color. We observe that the AM of VGT behaves as a monotonic function of $M_A$ in the sub-Alfvenic regime for both hydro-dynamical regimes. The $AM$ declines when $M_A$ increased. The $AM$ continues the declining trend throughout from sub-Alfvenic to trans-Alfvenic regimes. However, similar to the SIG result for supersonic ensembles, the $AM$ of VGT for supersonic ensembles becomes stable at about 0.4 at the transition from trans-Alfvenic to the super-Alfvenic regime,

A tendency of well alignment between VGT and magnetic field in the sub-sonic case is observed here. The AM of sub-sonic set always better than supersonic case with {\torefereetwo the AM improvement of about 0.2 throughout the change of $M_A$ from 0.2 to 1.2.}

\begin{figure*}
\label{fig:broadeningimshow}
\centering
\includegraphics[width=0.64\paperheight]{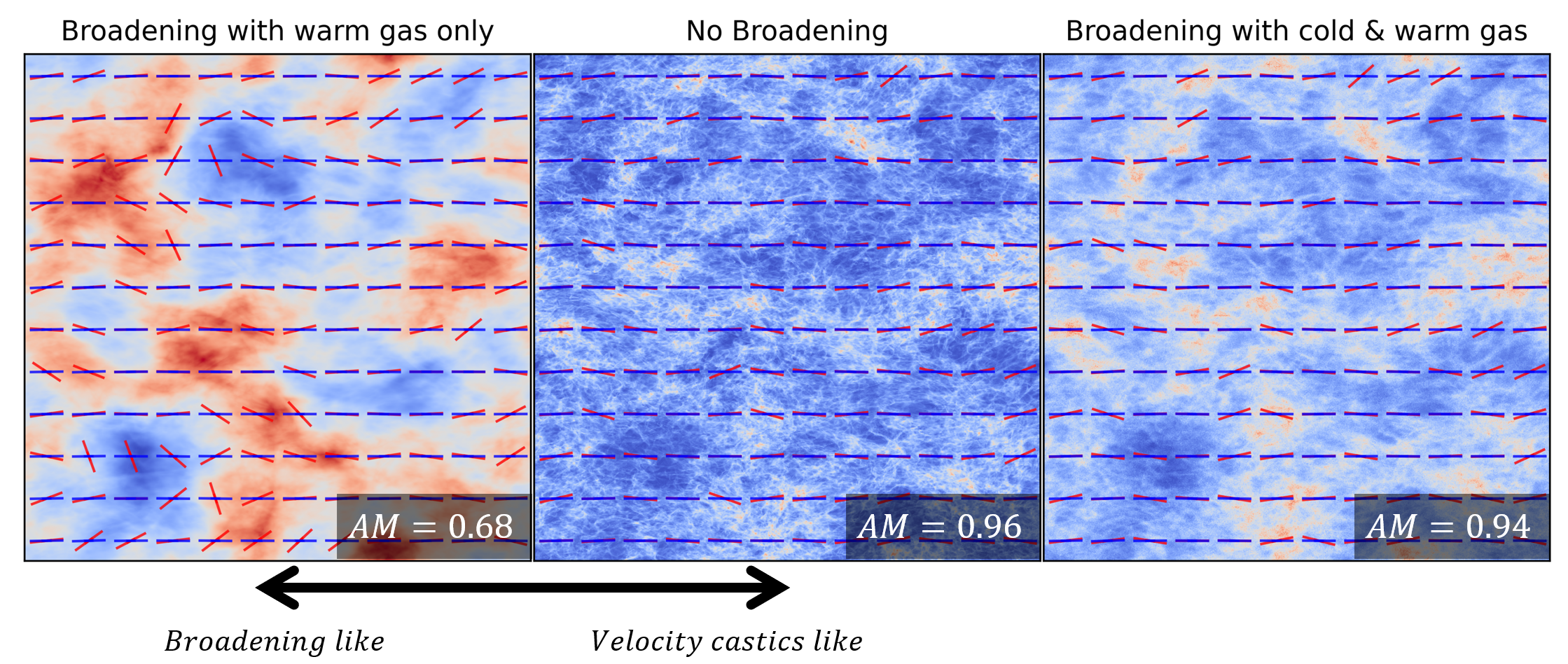}
\caption{{\toreferee The Comparison of intensity structure under the influence of thermal broadening. Simulation used in the figure: H4S. Warmer color means denser pixels and coolers means pixels with lower density. The blue and red arrow represents the magnetic field direction and gradient direction within the sub-block (block size = $66^2$). The bottom right shows the alignment measure value between magnetic field and gradient for each maps.}}
\end{figure*}

\subsection{Velocity Channel gradient in the multi-phase Interstellar medium}
\label{sec:Multiphase}

The Velocity Channel Gradients provide another way to study the magnetic field's morphology in the interstellar medium \cite{LY18a}.  The intensity fluctuation is strongly affected by its width and the thermal properties of the medium. \cite{HuNature} demonstrated the reliable performance of VChGs in tracing the magnetic field directions in  super-sonic molecular clouds. However, concerns of the thermal broadening effect were raised in a sub-sonic environment, which the effect could smooth out the velocity caustics in the channel maps {\toreferee \citep{Clark19}}. In the extreme case, {\toreferee when the thermal width larger than the velocity dispersion width}, the fine structure of the channel map would be washed out. In addition, this can makes it similar to the intensity map. However, the physics of the interstellar medium is complicated and involves external physical processes, especially for the HI medium. Thermal instability plays a crucial role in shaping the proprieties of the HI medium, resulting in the multi-phase interstellar medium. In multi-phase media, the numerical study found that the warm phase gas occupies most of the medium with about 5000K. On the other hand, the cold phase medium cools down to about 100K and occupied about $10\%$ space \citep{2003ApJ...586..1067H,2017NJPh...19f5003K,HO22A}.

Since two-phase media has a dramatic difference in temperature, the influence of broadening effect on the intensity structure in the channel map behaves entirely differently. The velocity profile of warm phase gas will greatly be extended because of its temperature and its fine structure in the channel map being affected. As a result, when we look at the transition of fine structure in channel map when switching different velocity channel, the caustics created in channel maps by turbulence in the warm phase gas will lose their contrast due to thermal broadening. A new technique, namely, the Velocity Decomposition Technique (VDA) can deal with the effect of thermal broadening and focus on the velocity caustics \citep{Yuen2021a}. In what follows, we another way of how the dynamics of warm gas can be revealed in the multi-phase medium.

If the multi-phase media is a unified turbulent system, dynamics between cold gas and warm gas are coupled \citep{spectrum}. The cold phase gas forms clumps that moving with the surrounding warm gas. It suggests the dynamical information of warm phase gas will imprint in the cold phase that is not much affected by thermal broadening. We expect this effect to be important in multi-phase galactic HI.  

{\toreferee To explore and verify this effect, we adopt a post-processing analysis to make synthetic observation of a multi-phase environment with broadening based on our sub-sonic ensembles simulation set. In our synthetic observation , we randomly select $15\%$ of pixels and label them as a cold phase gas tracer. We label the rest of the pixels as warm phase gas. We then transform the Position Position Position data cube (PPP) to Position Position Velocity (PPV) cube. We calculate a PPV cube accounting for a broadening effect. To do so, we {\toreferee convolved } each pixel with its temperature profile. To simplify our set up, we set the temperature of warm gas as 5000K and 100K for cold gas. The idea of the post-processing synthetic observation is inspired by \cite{Yuen2021a}. As noticed in \cite{LP00}, the fluctuation of channel maps can be divided into those arising from  density and velocity.  It is demonstrated in \cite{Yuen2021a} that, without changing the density value, one can vary the sound speed to change the fraction of density and velocity contributions in a channel map. We should stress that the isothermal simulation could not capture the full physics in multiphase ISM. However, \cite{Yuen2021a} demonstrated that the contribution of CNM and WNM in channel map could also be separated into the density and velocity part with the difference of different thermal profile.  As a result, we can apply two thermal profiles to the gas to try to simulate the behaviour of CNM and WNM in a channel map.}

Figure \ref{fig:broadeningimshow} demonstrates the center channel Map of synthetic observation from one of our simulation cubes(Right). As a reference, the figure also includes two comparison plots of the same Channel Map but one with a broadening effect with only warm phase (Left) and another one without broadening(Mid). This two picture represents two different regimes. In the sub-sonic regime, the morphology of the channel map without broadening shows a reference of intensity fluctuations caused by velocity caustics. Because of the existence of the velocity caustics effect, the channel map structure without broadening effect would demonstrate an intensity structure, which filling with thin and long filaments. Those intensity filaments caused by caustics within the thin channel map are elongated along the magnetic field, as described in LY18. On the contrary, the morphology of the channel map dominated by the broadening effect is different. In particular, the intensity fluctuation in the channel map is washed out because of the wide thermal velocity profiles. Therefore, the intensity structure in the channel map has a high similarity with the intensity maps. The similarity of the effects of thermal broadening and the increase of the thickness of the channel maps is discussed in LP00.

The situation is changed if we observe the intensity of emission in thin channel maps arising from the mixture of  warm and cold gas. There, the thin and long filamentary structures are clearly seen. This suggests that the main structure of velocity caustics is preserved in the the presence of multi-phase media with cold and warm gas mixed together.

\begin{figure}
\centering
\label{fig:Ch_AM_Broadening}
\includegraphics[width=0.49\textwidth]{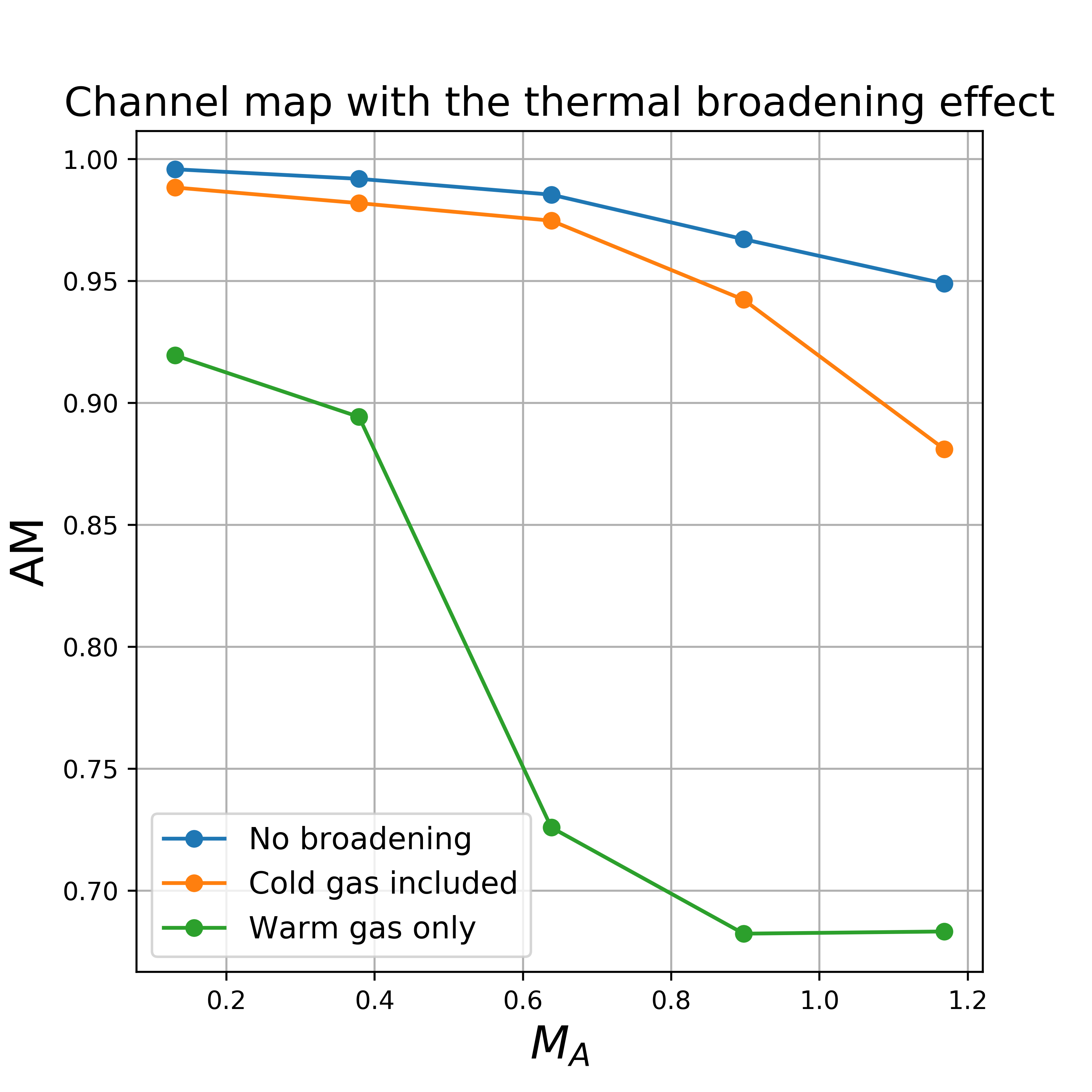}
\caption{Result of Channel Gradient considering the effect of thermal broadening. Block size used = 66. X-axis: Alfvenic Mach Number $M_A$, y-axis: AM }
\end{figure}

Figure \ref{fig:Ch_AM_Broadening} shows a scatter and line plot of AM of VGChT using channel map of multi-phase synthetic simulation with respect to $M_A$ using the gradient recipe same as the Figure \ref{fig:SC_AM}. The plot includes the $AM$ obtained in the channel maps with and without broadening. The AM for multi-phase simulation starting with $AM\sim 1.0$ in $M_A\sim 0.2$ with slowly decline to $AM\sim 0.88$ in $M_A\sim 1.2$. In contrast to the broadening regime, the AM curve for multi-phase simulation is very close to the velocity caustics regime in the sub-Alfvenic simulation with a small difference of AM. This discrepancy becomes broader as we transfer to the trans-Alfvenic environment.

\section{Improving AM in sub-sonic map using GGA technique}
\label{sec:GAG_imporvement}

\cite{HL20a} identified the effect of intermittency of fast mode in {\toreferee low-plasma $\beta$ media.}  Therefore, {\toreferee the concentration of fast modes in selected regions} would alter the anisotropy of the distribution of velocity centroids compared to the neighboring regions dominated by Alfvenic modes (see \cite{Kandel18}).

This effect would be reflected in the observed centroid gradients to abruptly change 90 degrees in the fast mode dominated regions.  We refer those gradients as orthogonal gradient. \cite{HL20a} introduced new data sets, namely, gradient amplitude map, and demonstrated that using these data sets one could suppress the orthogonal gradient effect.  As a result, the new gradient technique, Gradient of Gradient Amplitudes (hereafter GGA), could improve the alignment measure. The performance of GGA in ideal case (Environment without noise) could provide prefect alignment (AM$\sim$ 1) with the use of block size larger than $50^2$. 

\begin{figure}
\label{fig:GAG_Noise}
\includegraphics[width=0.49\textwidth]{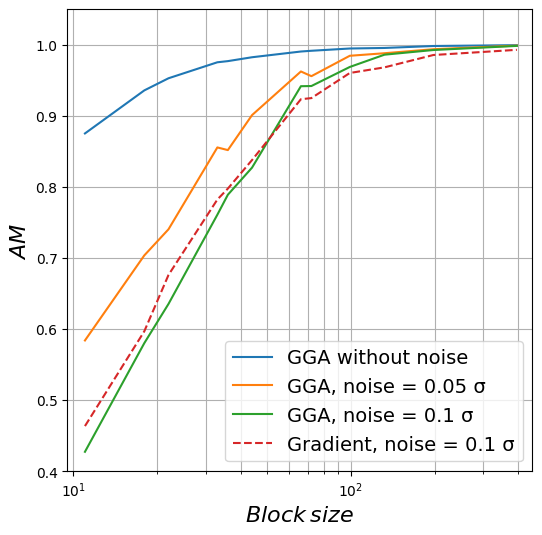}
\caption{The AM  of GGA versus the block size using synchrotron intensity. The line with different colors represent the performance of GGA with certain strength of white noise added. As a reference, the dotted line with red color illustrate the performance of gradient with the noise amplitude of $1 \sigma$. The x-axis showed in log scale for demonstrating the performance of technique in small block size.\\
Simulation used: H1S \\
Block size covered: [11,18,22,33,36,44,66,72,99,132,198,396] }
\end{figure}

However, we noticed that the performance of GGA could strongly depends on the level of noise. The performance of GGA will declin rapidly with the increase of noise. To demonstrate the effect of GGA in the presence of noise, we add white noise with the amplitude relative to the standard deviation of the observable measures and see how the $AM$ of GGA is varied as a function of noise amplitude. Figure \ref{fig:GAG_Noise} shows the $AM$ of GGA in centoid maps versus block size with white noise added of the amplitude 0.05 $\sigma$ and 0.1 $\sigma$. As a reference, we also added the AM of GGA without noise. Also, we include the AM of gradient with noise of 0.1 $\sigma$ for a comparison. 

{\toreferee For the computation of GGA, we first define the gradient amplitude map (GA), which mechanistically defined as $GA = \sum_i A_i^2$, where $A_i$ is gradient component in direction i. For the gradient technique, $A_i$ can be computed though the Sobel kernel. The GGA would then be the output of the Sobel kernel of GA.}

One can see from the figure, the performance of GGA drops rapidly with mild noise added. Compare to ideal case, the AM of GGA falls from $\sim 0.9$ to $\sim 0.6$ in the small block size For noise amplitude of $0.05 \sigma$. The performance gap narrows down with the larger block size but block size of $\geq 120^2$ is required to match the performance of ideal case. The advantage of GGA over ordinary gradient decreases for the case of noise amplitude  0.1 $\sigma$. We can see that the performance of GGA is very sensitive to the noise level if we use a smaller block size.

\begin{figure}
\label{fig:gag_smoothing_improvement}
\includegraphics[width=0.49\textwidth]{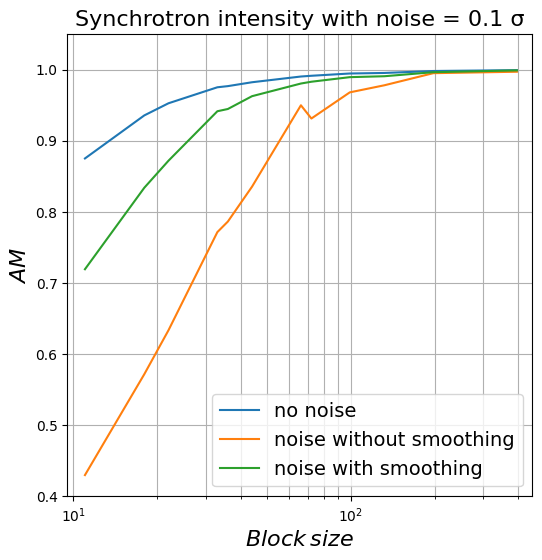}
\caption{The comparison of GGA before and after the smoothing technique using the synchrotron intensity map. As a reference, a blue line is added for representing the idea case.\\
Simulation used: H1S\\
Block size covered: [11,18,22,33,36,44,66,72,99,132,198,396] }
\end{figure}

To restore the performance of GGA, we employ the Gaussian smoothing of $\sigma = 2$ pixel as proposed in \cite{LY17a} and tested in \cite{LY18a}. According to \cite{LY17a}, the kernel size we picked here would preserve most of the small-scale structures while efficiently suppressing the noise in the synthetic map globally. By adding the noise and also the smoothing kernel, we can then test whether in noisy observations we can still use the GGA as a tool to trace magnetic field. Figure \ref{fig:gag_smoothing_improvement} shows the result of GGA verus block size with noise added of amplitude $0.1 \sigma$ and smoothing. The setup is the same as Figure \ref{fig:GAG_Noise}. We can see that the application of the smoothing technique shows that the performance of GGA can be improved. The drop of AM from 0.5 decrease to 0.8 in the small block size while the performance gap between smoothing and ideal case become negligible in the block size of $60^2$. The smoothing technique could relax the noise level requirement of the GGA.

\section{CFA in Gradient Amplitude Map}
\label{sec:CFA}
%Gs95->MHD

Other than gradient, Correlation Function Analysis(CFA) is another technique of tracing magnetic field direction by utilizing observable measure information \citep{EL05,Kandel17,2020ApJ...901...11H}. CFA was suggested to study magnetic field statistically and it is based on the theoretical understanding of properties of observed fluctuations (see LP12). {\toreferee For the (2 order) correlation function $CF_C$ of a velocity centroid map $C$, it is defined as
{\torefereetwo
\begin{equation}
\begin{aligned}
 CF_C({\bf R}) =  <C({\bf r})C({\bf r+R})>,
\end{aligned}
\end{equation}

where $r,R$ are the vector quantities }on 2D maps and separation distance from { \torefereetwo {\bf r}}. The output of 2D correlation map $CF_C({\bf R})$ can be interpreted as the fluctuations between different distance ${\bf R}$. If the fluctuations are isotropic, the shape of contour line will be circular. In opposite, the shape turns to elliptical when the fluctuation is anisotropic.} Therefore, the magnetic field direction could be obtained from the elongated direction of elliptical shape structure after the {\toreferee observational map} processed by the CFA analysis \citep{EL05}. The elongation depends on the relative importance of the three basic MHD modes in turbulence \citep{Kandel17}. It was applied to both observation and simulation data in \cite{Yuen18}. However, the study showed that the tracing power of the CFA is weaker and the technique is less stable than the gradient technique.  In this section, we  explore the behavior of CFA with the gradient amplitude maps.

A detailed study was conducted to compare the performance between gradient and other magnetic field tracing method, including CFA \citep{Yuen18}. One of the issues of CFA showed from \cite{Yuen18} is that the performance of CFA is not stable for the {\toreferee velocity centroid map}.  The anisotropy is changed when one selects a different block size (For example, figure 15 in \cite{Yuen18}). This change of anisotropy could change 90 degrees by switching the block size while the mean field's direction stays the same throughout the region. We repeated this study and extended it to the comparison between observable map and gradient amplitude processed map.

\begin{figure*}[t]
\label{fig:GA_CFA_illustration}
\centering
\includegraphics[width=0.64\paperheight]{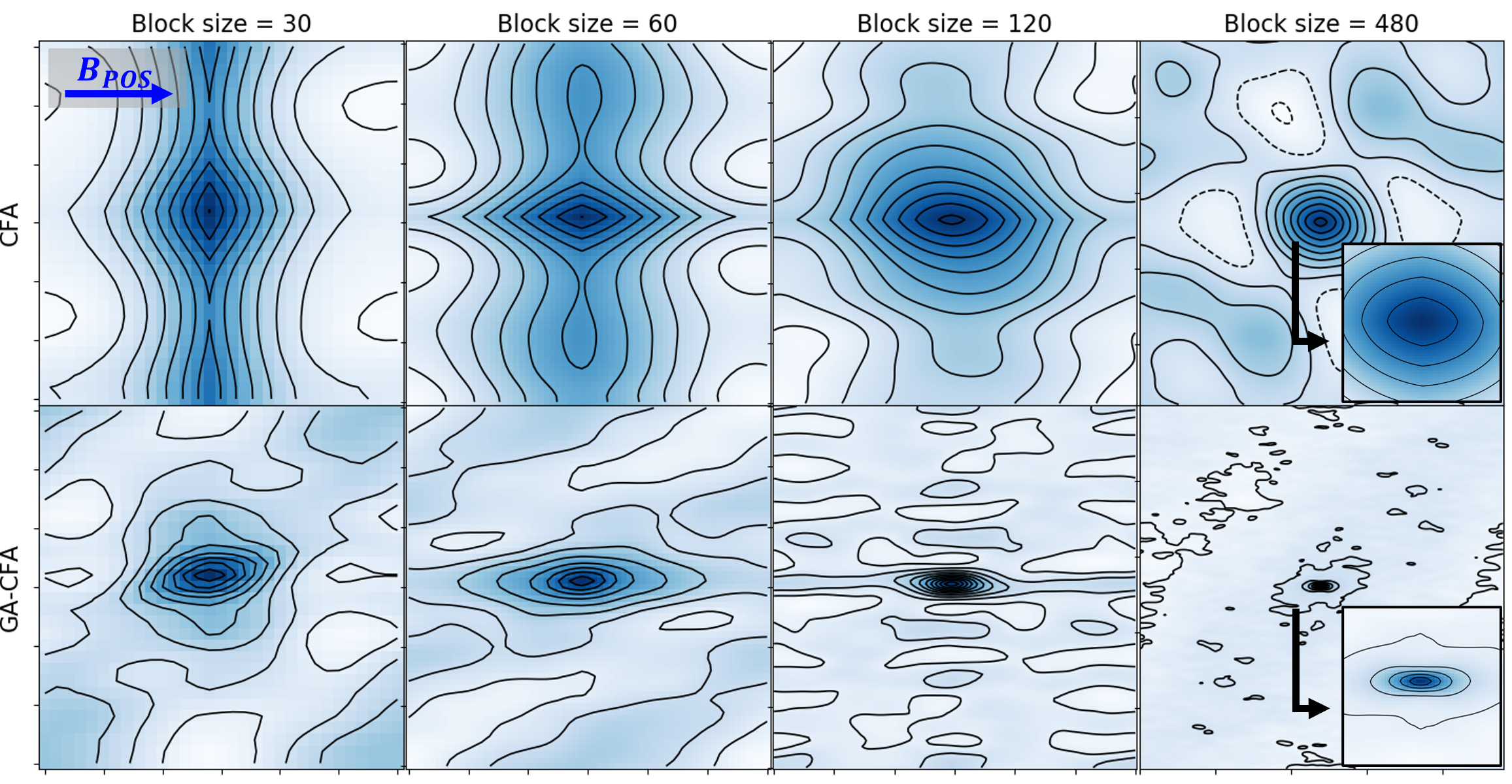}
\caption{Variation of correlation function anisotropy shapes with respect to block size. The subplot located at the right panel is the magnified view of the centre part of the plot. The blue arrow at the upper left plot shows the direction of the plane of sky magnetic field.\\
Top panel: CFA\\
Bottom panel: GA-CFA}
\end{figure*}

Figure \ref{fig:GA_CFA_illustration} shows how the shape of anisotropy of both maps is changed when one selects a different size of a averaging block. For sub-Alfvenic simulations like H3S, the mean magnetic field strength and direction remain the same throughout the region. For CFA, showed from the top side of the figure, we get the same conclusion as in  \cite{Yuen18}. While switching to the small size block region, the resolution problem can not only distort the shape of the anisotropies in different scales but also destroy the prominent elliptical shape. The shape of the elliptical structure is being destroyed for the block size is smaller than 120. Also, the direction of anisotropy changes when the block size changes. 

However, the situation improves dramatically with the application of the GA technique. For the procedure of processing GA-CFA, it is same as the computation of the CFA from \cite{Yuen18} but switching the input map to the gradient amplitude map. The bottom side of the figure shows the elliptical shape of CFA can be recovered after the GA technique. Nonetheless, the anisotropy stays towards horizontal direction throughout different block size. From the figure, {\torefereetwo We noticed that there are differences} between anisotropy direction and magnetic field in block size of $30^2$ but the anisotropy aligns with the magnetic field once increase the block size to $60^2$. On the other hand, one should mention that the size of the elliptical structure is smaller and more elongated compared to the normal CFA. The ellipse's shape exists on a small scale, about 20 to 60 pixels for GA-CFA, while it is about 40 to 60 pixels for the CFA. This is due to the map process after the gradient amplitude, the morphology of the map becomes more filamentary. The size of the filamentary structure is more prominent on a small scale  in the CFA analysis. So, to improve the tracing power of GA-CFA, we have to measure the direction of anisotropy on a smaller scale.

As the performance of CFA improved after combining with the GA technique, we then test the improvement of the new GA-CFA technique compared to gradient and GGA. We repeat the test showed in Figure \ref{fig:SC_AM} and extend it to both GGA and GA-CFA technique. {\toreferee Inspired by the result from figure \ref{fig:GAG_Noise} and figure \ref{fig:GA_CFA_illustration},  we observed a block size of $72^2$ would be a common "sweet spot" for both technique between the resolution required and the alignment improvement. We then pick the sub-block size of $72^2$ for the comparison.} The algorithm of determining the anisotropy direction of the CFA technique is the same as mentioned in the \cite{Yuen18}. {\toreferee For direct comparison with \cite{Yuen18}, we also adopt the same pixel distance of 10 pixels from the center of the elliptical structure for anisotropy contour detection.}

\begin{figure*}[t]
\label{fig:SC_AM_CFA}
\centering
\includegraphics[width=0.59\paperheight]{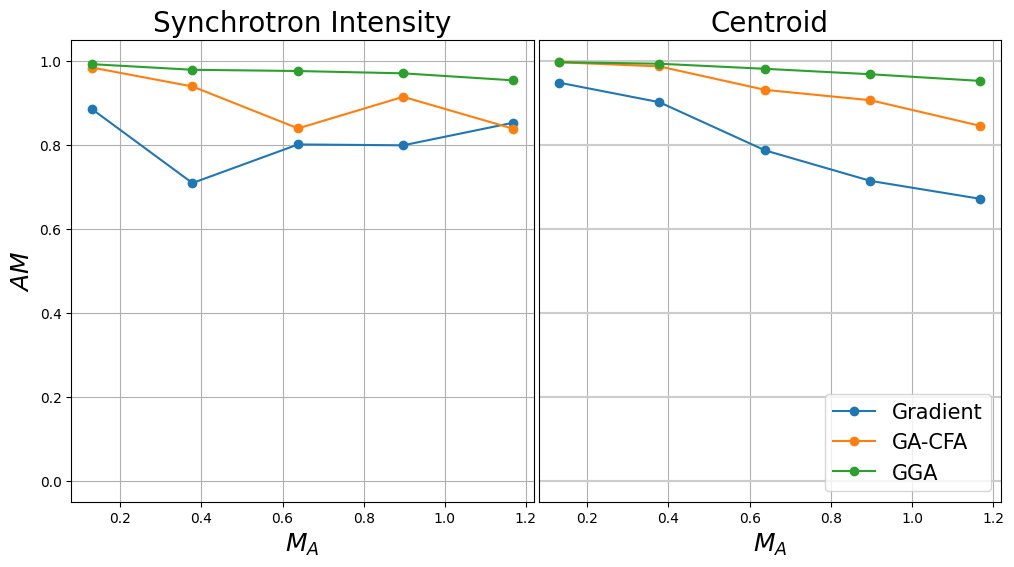}
\caption{Left panel: Result of Synchrotron intensity map betwen the gradient, GGA and GA-CFA. Right panel: Result of Centroid map betwen the gradient, GGA and GA-CFA. X-axis: Alfvenic Mach Number $M_A$, y-axis: AM.}
\end{figure*}

Figure \ref{fig:SC_AM_CFA} shows the results. One can see a significant advantage of GGA compared to the other two in the figure in terms of the AM. For the performance of GGA in both {\toreferee synthetic observation maps}, the AM decreases according to the Alfvenic Mach number. The performance drop is mild for GGA for the amount of $\Delta AM = AM_{M_A = 0.13}-AM_{M_A = 1.17} \sim 0.1$ when $M_A$ change from sub-Alfvenic to super-Alfvenic. The performance of the GA-CFA line between the gradient and GGA but closer to GGA in most of the cases but with a small effect of fluctuations. Compared to the gradient, GA-CFA has a noticeable better performance, which {\toreferee  AM improves by about 0.1 } for most cases.  This shows the performance of CFA can be improved by unitizing the Gradient amplitude technique. The synergy of the gradients and the GA-CFA approach will be explored elsewhere.

\section{Discussions} \label{sec:discussion}

\subsection{Connection to earlier gradient studies}
The gradient research  opens a new avenue of studying magnetic fields and turbulence properties and it is based on  of the modern understanding of MHD turbulence. Starting from the velocity centroids gradient in \cite{GL17}, studies employed later the gradient to different observable maps, such as synchrotron intensity/polarization \citep{LY17a}, channel maps \citep{LY18a}.  This enabled to trace the magnetic field in different  media from the molecular cloud on the scale of 0.1 pc to the galaxy clusters in the scale of 10kpc (see \cite{Hu20,Hu21}).  The applicability of gradient techniques covers two different hydrodynamics regimes to both sub-sonic to supersonic regimes. Meanwhile, the relationship between gradient and fundamental properties(such as $M_S, M_A$, and MHD modes) of MHD turbulence is being discovered. The gradient behavior could change 90 degrees in the particular region, for instance, shock or fast mode dominated region. In those regions, the direction of rotated gradient vectors would change from parallel to perpendicular to the magnetic field, which identifies as an orthogonal gradient region. Those orthogonal gradients could lower the tracing performance of gradient techniques.

This paper revisits  the performance of gradient techniques in the sub-sonic turbulent environment. We study the change of performance of different gradient techniques in sub-sonic medium regarding Alfvenic number systematically. In addition, we extend the study of gradient amplitude. This new technique showed a good performance in removing the distortion in the VGT-produced magnetic field maps arising from the effects of the intermittent regions dominated by fast MHD mode. This GGA technique was demonstrated to be capable of removing distortions caused by the fast mode. {\toreferee We noticed that GGA amplifies the small scale fluctuation that aligned with magnetic field, which suppresses the dominance of fast mode.} However, its performance is influenced by the noise and {\toreferee the reliability of GGA} could drop rapidly in the presence of the noise. We showed in section \ref{sec:GAG_imporvement} that the GGA performance in the presence of noise could be improved by employing suitable Gaussian filtering. This enables the new technique to be applied to realistic observation data. 

Furthermore, we explore a new way of combing the gradient amplitude maps with the CFA technique in section \ref{sec:CFA}. The classical CFA technique has its limitations while applied to the small block size region. This results in the requirement of block size $> 100^2$ and limits the abilities of the CFA in tracing magnetic field.  The new combined GA-CFA technique minimizes the block size e.g. to  $30^2$ without decreasing the performance of the technique.  This extends the applicability of the CFA technique and makes it competitive to the gradient technique. 

\subsection{Intensity structure and velocity caustics in Channel Map}

The theory of describing the fluctuations of intensity within spectroscopic data that arise from turbulence was  formulated in \citep{LP00}. There the concept of velocity caustics has been proposed to describe the effect of turbulent velocities eddies made to the channel map. This provided the basis for the technique of tracing magnetic fields using velocity channel gradients. 

However, the density also affect fluctuations in  channel map fluctuations.  Several HI studies have been discussed on the influence of thermal broadening of warm phase made to channel map and the importance of cold phase media. The applicability of \cite{LP00} to galactic HI was questioned in \citep{Clark19}. {\torefereetwo A rebuttal to these arguments was given by \cite{Yuen18}} and the applicability of the LP00-based approach was demonstrated in \cite{Yuen2021a} where the Velocity Decomposition Algorithm was introduced to deal with density fluctuations in subsonic flow. {\toreferee The later observational study by \cite{spectrum} reported the velocity caustics could be fully restored after applying the algorithm.}

Our study in section \ref{sec:Multiphase} showed provides another argument in favor of the applicability of the LP00 theory to multi-phase media. We showed that if the phases of the media move together in the galactic disk,  they can be viewed as a unified turbulent system, and our result from figure \ref{fig:Ch_AM_Broadening} suggests that most of the information of velocity anisotropy can be preserved without the VDA.  

\section{Summary} \label{sec:conclusion}
This paper extends our studies the Gradient Technique (GT) in the sub-sonic environment. Our main results are:

1. The alignment between gradient and POS magnetic field is better in the subsonic regimes compared to the supersonic one.

2. In the multi-phase media, the morphology of filamentary structure in the channel map and the statistical anisotropy of thin channel intensity fluctuations is preserved in the presence of thermal broadening if the phases are moving together. 

3. We extended the study of GGA introduced in the \cite{HL20a}. We examined the applicability of GGA in the synthetic observation map with noise added. The performance of GGA is sensitive to noise, but the employment of the Gaussian kernel alleviates the noise effect. 

4. We demonstrated that the gradient amplitude maps can be successfully combined with Correlation Function Analysis (CFA). In this case the anisotropy can is prominent in small block of the order of $30^2$. This makes the new technique competitive with the gradient technique.

\clearpage
\noindent
\section*{Acknowledgements}
We acknowledge Ka Ho Yuen and Yue Hu for the fruitful discussions. We acknowledge the support the NASA ATP AAH7546 and NASA TCAN 144AAG1967 grants.\linebreak
\noindent
\section*{Software}
Julia-v1.2.0/Julia-v1.8.2, Jupyter/miniconda3, LazTech-VGT \citep{YL17a} : https://github.com/kyuen2/LazTech-VGT
\section*{Data Availability}
The data underlying this article will be shared on reasonable request to the corresponding author.

\bibliographystyle{aasjournal}

\end{document}